# HUNTING FOR THE BEAUTY
# HEAVY-LIGHT QUARK SYSTEMS ON THE LATTICE[*]

STEPHAN GÜSKEN
*Physics Department, University of Wuppertal*
*D-42097 Wuppertal, Germany*

## ABSTRACT

We present selected topics on heavy-light physics on the lattice in order to illustrate the current status of this field. In particular results concerning $f_B$, semileptonic decays, the decay $B \to K^*\gamma$ and heavy baryon masses are discussed. Special emphasis is paid to the question of systematic effects which stem from the lattice discretization of QCD.

## 1. Introduction

The tremendous interest heavy-light quark systems have gained over the recent years is mostly due to the need of a precise determination of those Cabibbo Kobayashi Maskawa matrix elements, which govern the weak transition of heavy into light quarks. In order to extract these fundamental parameters from experimental measurements, an accurate non-perturbative calculation of the $QCD$ matrix elements involved is necessary. The aim of lattice $QCD$ in this context is to compute the required $QCD$ part directly from the very definition of the theory itself, i.e. it's action $S_{QCD}$.

The non-perturbative "per se" method to deduce $QCD$ properties directly from first principles would be "simply" to solve the $QCD$ path functional. For example the vacuum expectation value of an operator $O$ in terms of this functional is given by

$$\langle 0|O(\bar{\Psi},\Psi,A)|0\rangle = \frac{1}{Z}\int D[A]D[\bar{\Psi}]D[\Psi] O(\bar{\Psi},\Psi,A) e^{iS_{QCD}[g_0,A,\bar{\Psi},\Psi]} . \qquad (1)$$

Here $\Psi(\bar{\Psi}), A$ and $g_0$ denote the fermion fields, the gluons and the (bare) coupling. As it stands, however, Eq. (1) is mathematically ill defined. In order to give it a definite meaning one needs regularization.

The lattice method regularizes Eq. (1) by replacing the space-time continuum by an euclidian space-time lattice. In this way everything becomes well defined and the path functional can - in principle - be integrated out without any further approximation. In actual applications one mostly uses Monte Carlo techniques for this purpose.

---

Of course the lattice method has it's shortcomings. First of all discretization means introduction of a finite lattice spacing $a$, which corresponds to a finite energy cutoff $a^{-1}$. In order to extract continuum properties from lattice calculations, the influence of this cutoff onto the measured quantities must be investigated and finally an extrapolation $a \to 0$ is necessary. Clearly, on a given lattice, masses and energies have to stay away from the cutoff.
Secondly, a finite number of lattice points corresponds to a finite extension of the lattice in space and time. This, eventually, induces finite size effects, which also must be controlled and removed by an appropriate extrapolation. The great advantage of the lattice method is however, that it has *in itself* the power to control it's systematics: Lattice spacing and lattice size can be varied in order to visualize and finally to remove the corresponding lattice artefacts.

The investigation of heavy-light systems with lattice techniques is a very challenging task. On the one hand one is forced to keep the physical size of the lattice large in order to take care for the light – and thus long ranged – degrees of freedom. On the other hand the heavy degrees of freedom require a high lattice resolution, i.e. a small lattice spacing.
Current computer resources enable us to implement lattice resolutions up to $a^{-1} \simeq 2.5 - 3.5$ GeV, keeping finite size effects small. Thus, direct calculations in the $D$ meson region are feasible today. Beauty properties, however, can be accessed only by more elaborate lattice techniques. This *" hunting for the beauty"* is done with different, but interrelated methods.
The most direct attempt is to investigate the heavy quark mass dependence of the required quantities up to approximately two times the charm mass and then to extrapolate the results to the $B$ meson. This extrapolation can be replaced by a, clearly safer, interpolation if one performs in addition the lattice calculation at the heavy quark mass limit $m_h \to \infty$, where the heavy quark contribution can be integrated out.
A current, very active line of investigation addresses the question of how to reduce lattice artefacts by an improved discretization of the QCD action[1,2,3]. The general strategy is to include higher order (in $a$) terms into the discretized action, which weaken the cutoff dependence of the lattice data. Simulations using such an improved action are still "first principle" calculations, as the additional terms vanish in the continuum limit.
In a less fundamental approach one tries to reduce the lattice artefacts by a suitable redefinition of the quark field normalization[4,5,8]. Taking the free case as a guide, one successively employs mean field ideas and non-relativistic strategies in order to correct for the finite $a$ effects. The benefit of such an ansatz is still under debate, but clearly it always can be checked directly on the lattice by variation of $a$.

This talk is organized as follows. The next section will report on the status of $f_B$ on the lattice. Section 3 deals with the semileptonic decays $D \to K, K^*(\nu l)$ and $B \to \pi, \rho(\nu l)$. The decay $B \to K^* \gamma$ will be discussed in section 4. Section 5 is devoted to heavy baryon spectroscopy. Summary and conclusions will be given in section 6.[†]

All the lattice results presented here are obtained in the *quenched approximation*, which – loosely speaking – neglects internal fermion loops. It is used because of limited computer resources and – hopefully – it will be removed within the next years. Although it's impact cannot be quantified exactly at the moment, it has been proven to work accurately at least in the light fermion sector: quenched lattice calculations[6] of light fermion masses and decay constants are in agreement with the experimental results within an error margin of 10%.

## 2. Status of $f_B$

The decay constant $f_{PS}$ of a pseudoscalar particle is defined by the QCD matrix element
$$\langle 0|A_0|PS\rangle \equiv f_{PS} m_{PS} , \qquad (2)$$
with the 0-th component of the (heavy-light) axial current $A_0 = \bar{\Psi}_h \gamma_0 \gamma_5 \Psi_l$. It's discretized version $\langle 0|A_0|B\rangle^{latt}$, which can be evaluated on a given lattice, is related to the continuum matrix element by
$$\langle 0|A_0|PS\rangle = Z_A(g_0, m_h a, m_l a)\langle 0|A_0|PS\rangle^{latt} . \qquad (3)$$
The bare coupling $g_0$ and the quark masses $m_h$ and $m_l$ are the only free parameters on the lattice. The lattice spacing $a$ is determined in physical units by fixing one lattice quantity to it's experimental value. One frequently uses $m_\rho$, $f_\pi$ or $\sigma$(string tension) to set the scale.
The renormalization constant $Z_A$ is composed of a short distance part, which depends on the coupling $g_0$ and can be calculated in perturbation theory, and a long distance part, whose impact has to be removed by proper variation of $a$.

The question about the size of the systematic error introduced by the perturbative calculation of $Z_A$ is difficult to answer precisely. It has been shown over the recent years[7,8,10] that the replacement of the bare coupling $g_0$ by a suitably chosen effective coupling $g_{eff}$ leads to a clear improvement in the comparison of perturbative results and lattice simulations. Non-perturbative studies of $Z_A$ – including $ma$ effects – are also in progress[11].

In order to extract the matrix element $\langle 0|A_0|PS\rangle^{latt}$ one calculates the 2-point correlation function $\langle 0|A_0^\dagger(\vec{x},t)A_0(\vec{0},0)|0\rangle$, which can be decomposed into a sum over energy eigenstates $n$ with mass $m_n$
$$\sum_{\vec{x}} \langle 0|A_0^\dagger(\vec{x},t)A_0(\vec{0},0)|0\rangle = \sum_n |\langle 0|A_0|n\rangle|^2 e^{-m_n t} . \qquad (4)$$
The required ground state is obtained by analyzing the correlator at "suitably large" euclidian time $t$
$$|\langle 0|A_0|PS\rangle|^2 e^{-m_{ps} t} = \lim_{t\to\infty} \sum_n |\langle 0|A_0|n\rangle|^2 e^{-m_n t} . \qquad (5)$$

---
[†]We do not discuss here the results of non-relativistic lattice QCD. For a review the e.g. J. Sloan [9].

As the onset of the region of ground state dominance can be ascertained directly by analysis of the data, there is no basic problem to extract $\langle 0|A_0|PS\rangle^{latt}$.

On a finite lattice and with limited statistical accuracy one meets however serious difficulties in isolating the ground state unambiguously, especially in the case of heavy quarks[12,25]. The cure to this problem comes with the introduction of quark wave functions. Use of the exact ground state wave function would yield

$$\langle 0|A_0|n\rangle = 0 \quad \text{for } n \neq \text{groundstate} , \tag{6}$$

and $\langle 0|A_0|PS\rangle$ could be determined already at small values of $t$. An educated guess of the wave function equally leads to significant improvement, reminiscent to the variational approach in quantum mechanics.

*2.1. The Static Limit*

Considerable progress has been achieved in the determination of $f_B$ in the static limit with lattice methods during the last two years. The simulations have been improved with respect to their statistical significance – O(100) gauge configurations is standard already – and by the use of quark wave functions, which allow for a much more reliable ground state isolation. A variety of lattice results over a reasonable range of lattice spacings is available now such that the extrapolation to zero lattice spacing is feasible.

In Fig. 1 we show the collected results of the various groups.[‡] First of all we observe that the scaling combination $Z_A^{stat} f_{stat}^{latt} \sqrt{M_{PS}}$ exhibits a clear dependence on $a$ and an extrapolation to $a = 0$ is necessary. The PSI-Wuppertal collaboration[14] has performed this extrapolation on it's data, with the result $f_{stat} = 230 \pm 22 \pm 26$ MeV. This is nicely consistent with C.R. Allton's[22] result, $f_{stat} = 230 \pm 35$ MeV, who included all available data in the analysis.

A second look at Fig. 1 however reveals, that the data of the FNAL[15] and Kentucky[17] groups is significantly lower than the bulk of results at comparable $a$. This observation finds it's explanation in the fact, that these groups have developed and applied a new method of projecting onto the low energy eigenstates of Eq. 4. This "multistate smearing" method renders a conclusive test of ground state isolation and enables for the extraction of the mass gap between ground state and first excited state. The $a \to 0$ extrapolation of the FNAL data yields the result

$$f_{stat} = 188 \pm 23 \pm 15 \pm 26 \pm 14 \text{MeV} , \tag{7}$$

where the first error is due to statistics, the second gives the scale uncertainty, the third is due to the extrapolation in $a$ and the last one estimates the uncertainty of $Z_A^{stat}$. Although this result is still consistent with the former analysis mentioned above, it's somewhat lower value appears to be the more likely one[§]

---

[‡]The latest results of C.R. Allton et al.[19] are not discussed here , since a final analysis, including systematic uncertainties, is missing yet.

[§]R. Sommer showed[23] that the PSI-WUP data indeed moves down to the FNAL results if the information about the mass gap is included into the analysis. His re-analysis of the FNAL data, however, reveals a much larger uncertainty due to the extrapolation in $a$.

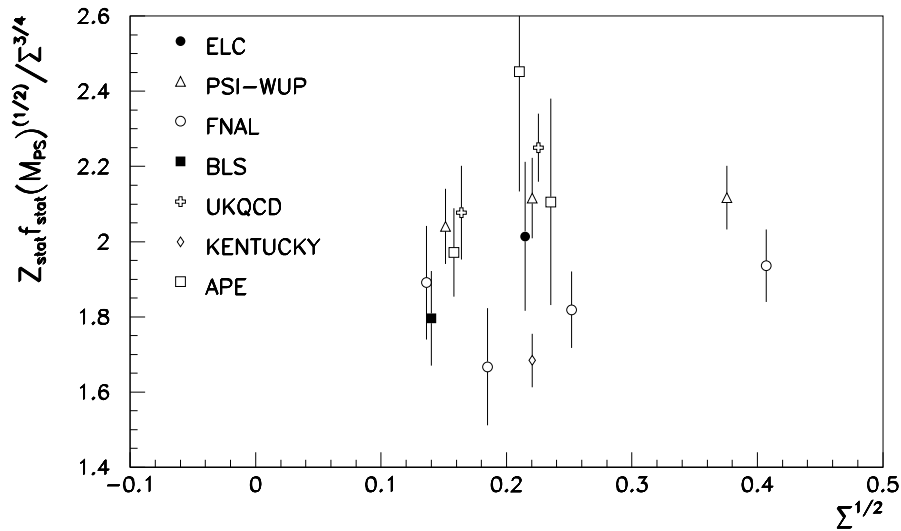

Figure 1: $Z_A^{stat} f_{stat}^{latt} \sqrt{M_{PS}}$ in units of the string tension $\Sigma = \sigma a$ as a function of $\Sigma$. The continuum is at $\Sigma = 0$. The data is taken from: ELC[13], PSI-WUP[14], FNAL[15], BLS[4], UKQCD[16], Kentucky[17] and APE[18]. In order to compare, all data have been uniformly rescaled with the string tension taken from Bali and Schilling[20] and with the mean field improved 1-loop[21] $Z_A^{stat}$.

### 2.2. $f_B$ with Propagating Quarks

Finite mass heavy-light systems on the lattice potentially bear the danger of being contaminated with large discretization errors, as one is tempted to work near the lattice cutoff, where the condition $aM_{PS} \ll 1$ is not valid.

In order to avoid this contamination one performs the lattice calculation away from the cutoff and then extrapolates¶ the results to the heavy mass region.

On top of this one tries to suppress the discretization errors by use of improved actions or redefinition of the quark field normalization, as explained in the introduction. A popular choice[16,18] of such an improved action is the Sheikholeslami-Wohlert action[1], which reduces the discretization effects from $O(a)$ to $O(a/lna)$, compared to the standard Wilson action[24]. Unfortunately one has to pay for it's application by larger statistical noise and an increased need of computer time and memory. Concerning the redefinition of quark field normalization[4,5,8] the question arises immediately, how it can be checked, that such ideas really do improve the situation. Fortunately the answer to that can also be given immediately: As improvement means to be – at given finite $a$ – closer to the continuum than in the unimproved case, one has to study the $a$ dependence of the observables in question. Weaker dependence on $a$ then signals improvement.

As an example of the feasibility of such test we show in Fig. 2 the results of the PSI-Wuppertal collaboration[25]. They have studied the decay constant $f_{PS}$ as a function of $a$ and $M_{PS}$, both in the standard Wilson normalization of quark fields and in the KroMac[5] normalization. Obviously the KroMac normalization does not weaken the $a$ dependence!

---

¶Or interpolates, if the result in the heavy mass limit is available.

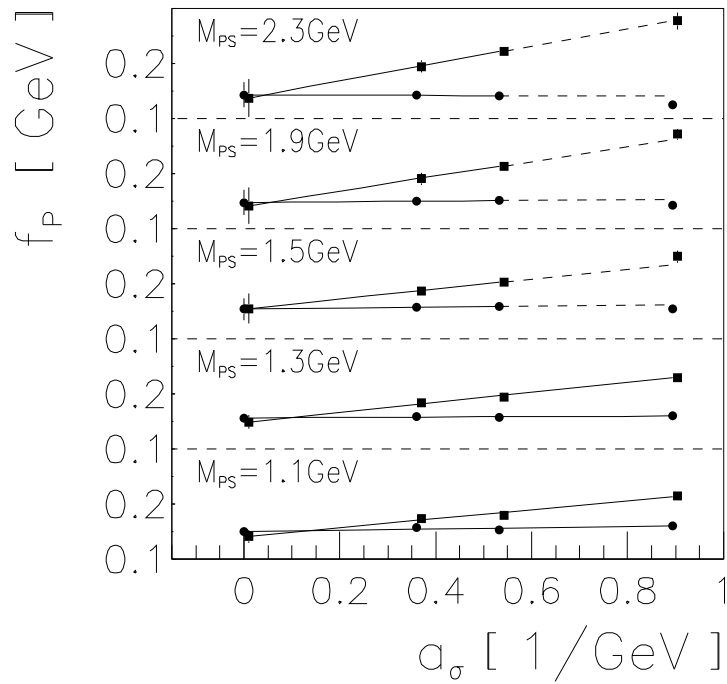

Figure 2: *The pseudoscalar decay constant $f_{PS}$ as a function of a, for various meson masses $M_{PS}$. The circles(squares) refer to the standard(KroMac) normalization. The solid lines indicate the extrapolation to the continuum. The continuum result is depicted at $a_\sigma = 0$ .*

Finally we present in table 1 the latest lattice results[||] concerning $f_D$ and $f_B$. It is encouraging to see that all results agree within errors, although different methods have been used in order to reduce finite $a$ effects. We emphasize that, for the first time, it has been possible really to investigate the finite $a$ and – which could not be discussed here – the finite size effects. A comparison with the results in the static limit (Eq. 7) furthermore reveals, that the large gap between $f_{stat}$ and $f_B$, which had been obtained in previous simulations[27], has almost completely disappeared. This is mostly due to (a) the extrapolation in $a$, (b) the use of proper quark wave functions and (c) the increased statistical accuracy of data.

## 3. Semileptonic Decays of Pseudoscalar Mesons

The next step is to calculate the QCD matrix elements relevant for the semilep-

---

[||]For comparison with results from QCD sum rules see e.g. C.A. Dominguez[26]

Table 1: *Latest results for $f_D$ and $f_B$ from various groups. $a_\sigma = \infty$ indicates that an extrapolation to the continuum limit has been performed. All groups have calculated both $f_{stat}$ and $f_{PS}$ with finite mass quarks.*

| group | $f_D$[MeV] | $f_B$[MeV] | method to reduce finite $a$ effects | $a_\sigma^{-1}$[GeV] |
|---|---|---|---|---|
| UKQCD[16] | $185^{+4+42}_{-3-7}$ | $160^{+6+53}_{-6-19}$ | SW action | $\simeq 2.7$ |
| BLS[4] | $208(9) \pm 35 \pm 12$ | $187(10) \pm 34 \pm 15$ | changed normalization | $\simeq 3.1$ |
| PSI-WUP[25] | $170(30)$ | $180(50)$ | cont. extrap. | $\infty$ |
| MILC[28] | $181(4)(18)$ | $147(6)(23)$ | cont. extrap. | $\infty$ |

tonic decays of pseudoscalar mesons. Compared to the calculations of $f_{PS}$ this is computationally more advanced, as one has to project onto *two* hadronic ground states, which asks for *two* large time separations on the lattice. Moreover, the need of spatial momentum $\vec{p} \neq 0$ states leads to an enhanced statistical noise of the lattice signals.

After the early pioneering work of C. Bernard et al.[33] and V. Lubicz et al.[38], much progress concerning the reliability of the results has been achieved over the last two years[29,30,31,34,35]. High statistics calculations (O(100) configurations), using quark wave functions and fine grained lattices are now available.

The $q^2$ dependence of the form factors, which parametrize the required matrix elements, has been studied in some detail. For $D$ decays all groups find consistency with the pole dominance hypothesis, though within still sizable statistical errors.

Conventionally, one extrapolates the form factors to their values at $q^2 = 0$. We emphasize that the validity of the pole dominance hypothesis in most cases is not crucial for the extrapolation, as the data contains already points near $q^2 = 0$. The corresponding results are collected in table 2.** We obtain, that all estimates are in reasonable agreement with experiment.

One might ask whether the existing lattice data allow already for an extrapolation $a \to 0$. In Fig. 3 we investigate this question for the case of $f_D^+(0)$. It is encouraging to see that the data, covering already a sizable range in $a$, exhibit only a weak (if any) dependence on the lattice spacing. Due to the large errors however, an extrapolation in $a$ appears to be premature.

In order to obtain the form factors for the decays $B \to \pi$ and $B \to \rho$, one has to extrapolate in the heavy meson mass. In view of the accuracy of the existing data, this is a difficult task, but ELC[30], APE[31] and UKQCD[32] have tackled it. Their findings are displayed in table 3, together with a comparison to QCD sum rules. Clearly, the quality of data has to be increased in order to make accurate predictions.

---
**This is a abbreviated version of the nice compilation given by the UKQCD[29] group. A comparison with QCD sum rules and potential model calculations is also given there.

Table 2: *Form Factors at $q^2 = 0$ for the semileptonic decays $D \to K, K^*$.*

| group | $f_K^+(0)$ | $f_K^0(0)$ | $V(0)$ | $A_1(0)$ | $A_2(0)$ |
|---|---|---|---|---|---|
| UKQCD[29] | $0.67^{+7}_{-8}$ | $0.65(7)$ | $1.01^{+30}_{-13}$ | $0.70^{+7}_{-10}$ | $0.66^{+10}_{-15}$ |
| ELC[30] | $0.60(15)(7)$ | | $0.86(24)$ | $0.64(16)$ | $0.40(28)(4)$ |
| APE[31] | $0.72(9)$ | | $1.0(2)$ | $0.64(11)$ | $0.46(34)$ |
| BKS[33] | $0.90(8)(21)$ | $0.70(8)(24)$ | $1.43^{+45+48}_{-45-49}$ | $0.83(14)(28)$ | $0.59^{+14+24}_{-14-23}$ |
| BG[34] | $0.73(5)$ | $0.73(4)$ | $1.24(8)$ | $0.66(3)$ | $0.42(17)$ |
| WUP[35] | $0.76(15)$ | $0.75(6)$ | $1.05(33)$ | $0.59(8)$ | $0.56(40)$ |
| LMMS[38] | $0.63(8)$ | | $0.86(10)$ | $0.53(3)$ | $0.19(21)$ |
| Experiment(a)[36] | $0.77(4)$ | | $1.16(16)$ | $0.61(5)$ | $0.45(9)$ |
| Experiment(b)[37] | $0.70(3)$ | | | | |

Table 3: *Form Factors at $q^2 = 0$ for the semileptonic decays $B \to \pi, \rho$. The subscripts "naive", "HQET" refer to the way in which the extrapolation in $m_{PS}$ was done.*

| group | $f^+(0)$ | $V(0)$ | $A_1(0)$ | $A_2(0)$ |
|---|---|---|---|---|
| APE$_{HQET}$[31] | $0.29(6)$ | $0.45(22)$ | $0.29(16)$ | $0.24(56)$ |
| APE$_{naive}$[31] | $0.35(8)$ | $0.53(31)$ | $0.24(12)$ | $0.27(80)$ |
| ELC$_{HQET}$[30] | $0.26(12)(4)$ | $0.34(10)$ | $0.25(6)$ | $0.38(18)(4)$ |
| ELC$_{naive}$[30] | $0.30(14)(5)$ | $0.37(11)$ | $0.22(5)$ | $0.49(21)(5)$ |
| UKQCD[32] | $0.24^{+0.04}_{-0.03}$ | | | |
| Ball[39] | $0.26(2)$ | $0.6(2)$ | $0.5(1)$ | $0.4(2)$ |

**4. The Decay $B \to K^* \gamma$**

To leading order perturbation theory in the weak coupling, the matrix element for the decay $B \to K^* \gamma$ is given by[40]

$$M = \frac{e G_F m_b}{2\sqrt{2}\pi^2} C_7(m_b) V_{tb} V_{ts}^* \langle K^* | J_\mu | B \rangle \; , \qquad (8)$$

where $J_\mu = \bar{s}\sigma_{\mu\nu}q^\nu(1+\gamma_5)b$, and $\eta$ and $q$ are the polarization and momentum of the emitted photon. The non perturbative QCD matrix element $\langle K^* | J_\mu | B \rangle$ can be parametrized by three form factors[41] $T_i(q^2)$, $i=1,2,3$, which need to be evaluated at $q^2 = 0$, since the emitted photon is on-shell. In this limit the form factors obey the relations

$$T_2(q^2 = 0) = -iT_1(q^2 = 0) \qquad T_3(q^2 = 0) = 0 \; . \qquad (9)$$

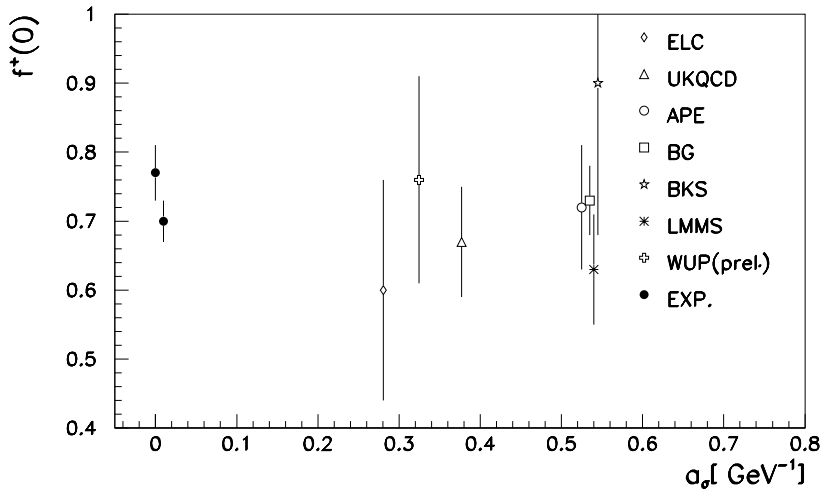

Figure 3: $f_D^+(0)$ as a function of $a$. The experimental results are shown at $a = 0$. References are found in Tab. 2. For clearness the data at $a_\sigma \simeq 0.53\ GeV^{-1}$ have been shifted slightly.

Clearly, the task of lattice QCD is to determine $T_1(q^2 = 0)$ and – for consistency – $T_2(q^2 = 0)$. With current lattice capabilities this requires extrapolation to $M_B$ as well as to $q^2 = 0$.

As the functional dependence of $T_1(q^2, M_{PS})$ and $T_2(q^2, M_{PS})$ is not a priori known, one has work with ansätze, motivated by pole dominance models and heavy quark symmetry. This is not a problem in principle, because the reliability of these assumptions can be tested directly on the lattice. Of course the final evidence may then depend crucially on the accuracy of data. Three groups[42,43,44] have recently studied the decay $B \to K^*\gamma$. The most detailed analysis has been performed by the UKQCD collaboration[44]. They find

$$T_1(q^2 = 0) = \begin{cases} 0.159^{+34}_{-33} \pm 0.067\ (a) \\ \\ 0.124^{+20}_{-18} \pm 0.022\ (b) \end{cases}, \qquad (10)$$

using the assumption of (a) single pole dominance and (b) double pole dominance for the $q^2$ dependence of this form factor. The corresponding value of $T_2(q^2 = 0)$ is consistent with these results if single pole dominance is assumed for $T_2$. Unfortunately the data is not accurate enough to proof the assumptions.

The results of Bernard et al.[42] and Abada et al.[43] are consistent (within large errors) with the values quoted above. However, the latter ascertain a crucial influence of the assumptions made for the $q^2$ dependence. Thus, much higher accuracy of data is needed in order to extract the functional dependence directly from lattice simulations.

### 5. Heavy Baryon Spectroscopy

The determination of baryon masses with lattice methods is a much easier and clearer task than the evaluation of form factors. They are extracted from the time decay of 2-point correlation functions, which exhibit a comfortably high signal to noise ratio, at spatial momentum $\vec{p} = \vec{0}$. Moreover one does not have to struggle with perturbative renormalization constants, because baryon masses are renormalization group invariants. Correspondingly, reliable extrapolations to the $b$ quark mass and to $a = 0$ are feasible with today's capacities. In Fig. 4 we display the mass difference $M_\Lambda - M_{PS}$ as a function of $1/M_{PS}$ at several lattice spacings $a_\sigma$, as found by the PSI-Wuppertal[45] collaboration. We observe that it behaves nicely linear $1/M_{PS}$. The finite $a$ effects turn out to be small, which allows for a smooth extrapolation in $a$. They finally get

$$M_{\Lambda_b} = 5.728 \pm 0.144 \pm 0.018 \text{ GeV} ,\qquad(11)$$

which is consistent with the result of the UKQCD collaboration [46]††

$$M_{\Lambda_b} = 5.900^{+0.170}_{-0.150} \text{ GeV} ,\qquad(12)$$

obtained on a single lattice with $a_\sigma \simeq 2.7 \text{GeV}$.

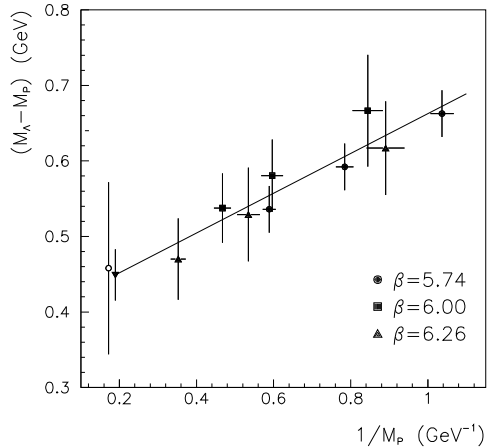

Figure 4: $M_\Lambda - M_{PS}$ as a function of $1/M_{PS}$ and $a$. The solid line corresponds to a global fit assuming no $a$ effects. It gives the value shown with the inverted triangle at the B meson mass. The value obtained after extrapolation to the continuum is indicated by the open circle. $\beta = 5.74, 6.00, 6.26$ corresponds to $a_\sigma^{-1} \simeq 1.12, 1.88, 2.78\, GeV$.

## 6. Summary and Conclusions

In this talk we described the systematic improvements achieved recently on computing heavy-light physics on the lattice, and presented the latest results from such simulations. We have seen that – although the large mass of the beauty quark still impedes a direct evaluation – high precision calculations in the range up to two times the $D$ mass allow already in some cases for a safe extrapolation to the $B$ meson. Due to the advent of fast parallel computers the cutoff dependence of lattice results can be studied now in detail, which makes the extrapolation to the continuum feasible. Given this necessary condition, the door is open to all kinds of improvement ideas, whose benefit can be tested with the "lattice lab".

---

††A comprehensive list of UKQCD results concerning beautiful Baryon masses is given also in that reference.

Current lattice capabilities have enabled us to consolidate the value $f_B = 180(50)$MeV and to give reliable estimates for the form factors of semileptonic $D$ decays. First high statistics results concerning the decays $B \to \pi, \rho$, $B \to K^*\gamma$ and the masses of beautiful baryons are also available now.

One might wonder, whether the "hunting for the beauty" will end up in a "rendezvous". In the worst scenario, the hunter will have to wait three years, to the advent of 60 Gflops machines, which are powerful enough to increase lattice resolution from $a^{-1} \simeq 3$GeV to $a^{-1} \simeq 6$GeV. But there are many good ideas to reach the goal in a shorter time.

## 7. Acknowledgements

I thank Hugh P. Shanahan for providing me with the latest UKQCD results concerning the decay $B \to K^*\gamma$.